\documentclass[12pt]{article}
\title{
BCS-Bose Crossover in Color Superconductivity
\\}
\author{B.O. Kerbikov\\ State Research
Center\\Institute of Theoretical and Experimental Physics, \\
Moscow, Russia}
 \date{}
  \newcommand{\be}{\begin{equation}}
\newcommand{\ee}{\end{equation}}
\def\la{\mathrel{\mathpalette\fun
<}} 
\def\fun#1#2{\lower3.6pt\vbox{\baselineskip0pt\lineskip.9pt
\ialign{$\mathsurround=0pt#1\hfil ##\hfil$\crcr#2\crcr\sim\crcr}}}

\newcommand{\ver}{\mbox{\boldmath${\rm r}$}}

\begin{document}
\maketitle

\begin{abstract}
It is shown that the onset of the color superconducting phase
occurs in the BCS-BE crossover region.
\end{abstract}

During the last 2-3 years color superconductivity became one of
the QCD focal points --see review papers \cite{1}-\cite{3}. To a
large extent (but not completely) the basic ideas of the subject
are traced back to the BCS theory of superconductivity and its
later development. It is known that the discovery of high
temperature superconductors (HTSC) gave rise to new ideas and
approaches and revealed interest to the problem of the transition
from the BCS regime to the Bose-Einstein (BE) condensation. The
BCS-BE crossover is important for the physics of HTSC since the
underlying distinction of HTSC from ordinary superconductors is
that they are characterized by much smaller value of the
dimensionless parameter $\xi n^{1/3}$, where $\xi$ is the
coherence length and $n$ is the carrier density. In the BCS, or
weak coupling regime, $\xi^3n\sim(10^8-10^{10})$ while in the
opposite strong coupling case $\xi^3n\la 1$ and we are dealing
with the compact pairs of composite bosons, which may undergo BE
condensation. It has been suggested (see e.g. \cite{4}) that the
description of the HTSC might require an intermediate approach
between the BCS and BE limits. The evolution from weak to strong
coupling was theoretically investigated \cite{5} before the
discovery of HTSC. It was shown that the transition proceeds via a
smooth  crossover though the two limits are physically quite
different (see also \cite{6}).

Having reminded these well known facts we may formulate the
question which forms the core of the present note. As model
calculations show \cite{7} the onset of the color superconducting
phase in two flavor QCD(the so called 2SC phase) occurs at rather
low quark densities $n$, namely at $n$ only three times larger
than the quark density in normal nuclear matter, or even at lower
ones \cite{8}.

Thus it is natural to ask in which region regarding the BCS-BE
crossover  does it happen?

According to \cite{7} in  QCD with two massless flavors transition
to the superconducting $2SC$ phase occurs at $n^{1/3}\simeq
0.2$GeV (the dimension of $n$ is $1/fm^3$ or GeV$^3$). As for the
corresponding value of $\xi$ we may only rely on some estimates
since accurate calculations are lacking. One should also keep in
mind possible distinction between the correlation length and the
pair size. The two quantities coincide in the BCS regime  \cite{9}
while in the BE region the pair size is smaller than the coherence
length \cite{10,6}. With these reservations being made we quote
the value $\xi\simeq 0.8$ fm from \cite{11}. Rather close result,
namely $\xi \simeq 0.6$ fm, follows from the BCS estimate $\xi
\simeq 1/\pi\Delta$ \cite{12}, where $\Delta\simeq 0.1 $ GeV
\cite{1,7}. Thus the diquark pair in the "newly born" color
superconducting phase is rather compact. This is easy to
understand from simple physical considerations. In color
antitriplet  $\overline 3$ state the one-gluon exchange leads to
the quark-quark potential which is only a factor of two weaker
than the quark-antiquark one. Instanton or NJL models also result
in a rather strong $q-q$ attraction. Consider the NJL "weak
coupling" solution \cite{1,3} $\Delta=2\omega_D \exp
(-1/\rho_{NJL}$), where $\omega_D$ is the Debye frequency,
$\rho_{NJL} =8g^2\mu^2/\pi^2$, $g^2\simeq 2 GeV^2$ is the NJL
coupling constant, $\mu\simeq 0.4 GeV$ is the chemical potential
corresponding to the onset of the 2SC phase. We immediately see
that $\rho_{NJL} \simeq 0.3 > \rho_{BCS}$, i.e. the quark-quark
interaction is  stronger than phonon mediated electron-electron
interaction and in this sense the "newly born" color
superconducting phase does not correspond to the standard BCS weak
coupling limit.

We conclude that the onset of the color superconducting phase
corresponds to $\xi n^{1/3}\sim (1fm) (0.2 GeV) \sim 1, ~~\xi \mu
\sim (fm) (0.4 GeV) \sim 2$.

These values are at least two orders of magnitude smaller than
those corresponding to the BCS regime. In order to understand to
which region of the BCS-BE "phase diagram" they correspond one has
to resort to model calculations performed for the system of
electrons. Most results have been obtained for the system in two
dimensions \cite{6}. Crossover in three dimensions has been
studied in \cite{13}. Transition between the two regimes occurs in
a narrow range of the parameter $\xi n^{1/3}$ and for the electron
systems with simple model potentials (finite range, separable,
Gaussian) the value $\xi n^{1/3}\simeq 1$ corresponds to the lower
limit of the BCS-like region. Needless to say that the
extrapolation of this result to the system of massless quarks may
be considered only as an educated guess and the problem deserves a
dedicated study.

The importance of the BCS-BE  crossover for  the color
superconductivity problem was first outlined on \cite{3} and
\cite{14a} -\cite{15a}. In \cite{14a} the low density regime was
investigated by extrapolating the single-gluon exchange model
which is an adequate tool at asymptotically high densities. The
smooth transition from $\xi n^{1/3}\gg 1$ to $\xi n^{1/3} =10$ at
$\mu=0.8$ GeV was observed.

 Let us indicate
  how the crossover
problem in color superconductivity theory has to be approached.
The starting point in the thermodynamic potential $\Omega(\Delta,
\mu, T)$, where the diquark condensate $\Delta $ is of the form
$\Delta \propto<q\hat 0 q>,$ $ \hat 0=\varepsilon_{\alpha\beta3}
\varepsilon_{ij} C\gamma_5,~~ \alpha,\beta$ -- are the color
indices, $i,j$- flavor ones, $C$- change conjugation operator
\cite{1}. The dependence of $\Omega$ on the chiral condensate
$\varphi\propto <\bar q\hat R q>,~~ R=\varepsilon_{\alpha\beta}
\varepsilon_{ij}$, in the region where the two condensates
$\Delta$ and $\varphi$ possibly coexist \cite{7} may be dropped
due to the color superconductivity version of the Anderson theorem
\cite{14}. For a wide class of models with four-quark interaction
the  thermodynamic potential $\Omega$ was first analytically
calculated in \cite{7}. With $\Omega$ at hands one can write
self-constituent set of mean-field equations to determine the gap
$\Delta $ and the chemical potential $\mu$: \be
\frac{\partial\Omega}{\partial\Delta}=0,~~
-\frac{\partial\Omega}{\partial\mu}=n. \label{1} \ee

Up to now only the first one of equations (\ref{1}) has been used
in color superconductivity theory while the chemical potential has
been considered as independent variable. To display the crossover
from the BCS to the low density BE regime one also needs the
second equation \cite{6}. It enables to consider the region of
$\mu<0$ values characteristic for the delute gas of tightly bound
diquarks $(\mu=-\varepsilon_{B/2}$ in the limit of delute
composite bosons with binding energy $\varepsilon_B$).

In order to find the quantities $\Delta $ and $\mu$  vs the
dimensionless physical parameter $\xi n^{1/3}$ equations (\ref{1})
have to be complemented by the equation which determines the
parameter $\xi$ \be
 \xi^2=\frac{\int d\ver \phi (r) r^2}{\int d\ver \phi
 (r)},\label{2}
 \ee
 where $\phi(r) =<q(r) \hat 0 q(0)>$. Analytic expressions for
 $\Delta$ and $\mu$ as functions of $\xi n^{1/3} $ have been
 obtained for electron systems with simple model interaction
 \cite{13}. The necessaty to implement similar program for color
 superconductivity directly follows from our conclusion that
 the suggested onset of this phase occurs within the BCS-BE
 crossover region.

 Another question is what are the physical consequences of the
 fact that the formation of the color superconducting gap is at
 least partly due to the existence of the preformed Bose pairs of
 quarks. At present we can again rely only on the corresponding
 studies of the electron systems \cite{5,15}. The key point here
 is that the physical origin of the critical temperature $T_c$ is
 absolutely different in the limits of weak and strong coupling
 \cite{5}. In the BCS region $T_c$ corresponds to the breaking of
 Cooper pairs while in the BE limit $T_c$ corresponds to the pairs
 center-of mass motion and to the population of zero-momenta
 state. Transition from the weak to strong coupling regimes
 results in the decrease of $T_c$. comparing to mean-field value.
  Formally this should also
 follow from equations (\ref{1})-(\ref{2}).

 Finally we note that calculations of the parameters $\mu$ and $T$
 at which the  transition into color superconducting phase occurs
 have been performed  neglecting the gluon condensate. General
 arguments presented in \cite{16} show that color-magnetic
  field which is "frozen"
 into the quark system in the form of the gluon condensate shifts
 the transition towards higher densities. (see also \cite{17}).
 Therefore equations (\ref{1}) for the BCS-BE crossover should be
 embedded into the background gluon field.

 The author  is thankful to V.I.Shevchenko and Dirk van der Marel
 for useful discussion and remarks
  and to Kazunori Itakura and Egor Babaev for bringing attention
  to Refs. \cite{14a} and \cite{15a}.  The author acknowledges receipt
 of financial support from the grants RFFI-00-02-17836, INTAS-110
 and NWO-01-250.

\end{document}